# The Design of a 3D Character Animation System for Digital Twins in the Metaverse

**Senem Tanberk[1], Dilek Bilgin Tukel[2,3], and Kadir Acar[3]**

[1]Research and Innovation, Huawei Türkiye Research and Development Center, Istanbul, Türkiye
[2]Department of Software Engineering, Dogus University, Istanbul, Türkiye
[3]Research and Development, Altinay Robot Technologies, Istanbul, Türkiye

Corresponding author: Senem Tanberk (e-mail: senem.a.tanberk@gmail.com)

**ABSTRACT** In the context of Industry 4.0, digital twin technology has emerged with rapid advancements as a powerful tool for visualizing and analyzing industrial assets. This technology has attracted considerable interest from researchers across diverse domains such as manufacturing, security, transportation, and gaming. The metaverse has emerged as a significant enabler in these domains, facilitating the integration of various technologies to create virtual replicas of physical assets. The utilization of 3D character animation, often referred to as avatars, is crucial for implementing the metaverse. Traditionally, costly motion capture technologies are employed for creating a realistic avatar system. To meet the needs of this evolving landscape, we have developed a modular framework tailored for asset digital twins as a more affordable alternative. This framework offers flexibility for the independent customization of individual system components. To validate our approach, we employ the English peg solitaire game as a use case, generating a solution tree using the breadth-first search algorithm. The results encompass both qualitative and quantitative findings of a data-driven 3D animation system utilizing motion primitives. The presented methodologies and infrastructure are adaptable and modular, making them applicable to asset digital twins across diverse business contexts. This case study lays the groundwork for pilot applications and can be tailored for education, health, or Industry 4.0 material development.

**INDEX TERMS** Digital twin, metaverse, virtual simulation, avatar animation, breadth-first search, peg solitaire, board game.

## I. INTRODUCTION

The metaverse represents a new era characterized by an immersive digital realm, social connectivity, and three-dimensional (3D) experiences, breaking the boundaries between the physical and virtual worlds. As this digital domain continues to evolve, the need for robust infrastructure likened to Web 3.0 necessitates the integration of various technologies such as artificial intelligence (AI), augmented reality (AR), virtual reality (VR), the fifth generation (5G) and sixth generation (6G) of wireless technology, cloud computing, edge computing, mobile applications, and blockchain [13, 18, 24, 25, 41]. One key aspect of this infrastructure is the creation of asset digital twins that are virtual replicas of real-world objects, environments, or entities, seamlessly integrated with dynamic three-dimensional (3D) animations. These digital twins serve as the building blocks of the metaverse, facilitating immersive experiences across a wide range of use cases including manufacturing, education, entertainment, collaboration, and commerce [3, 4, 10, 11].

The concept of asset digital twins offers unparalleled levels of immersion and interactivity with digital content by combining 3D models with realistic animations within the metaverse. Asset digital twins enable users to engage with virtual environments in ways that closely mimic real-world experiences, whether it involves exploring a virtual museum, participating in a multiplayer game, analyzing industrial automation, or attending a virtual conference.

The development of asset digital twin infrastructure for the metaverse presents both technical and creative challenges. On the technical front, considerations such as data storage, retrieval, and streaming bandwidth must be addressed to run a system entailing 3D content smoothly across different platforms and devices. Concurrently, in the metaverse, creative endeavors are required to design aesthetic 3D models and animations that accurately capture and visualize real-world objects and environments.

Collaboration and interdisciplinary approaches are essential to the success of asset digital twin infrastructure development. For a holistic approach to innovation, it is necessary to bring together experts from fields such as 3D modeling, animation, software engineering, user experience design, and domain-specific industries to work together on a project-specific basis. By fostering collaboration and the sharing of knowledge, organizations can harness the collective expertise of diverse teams to overcome challenges in metaverse development. In the industrial context, the metaverse offers novel opportunities for collaboration, automation, and remote operation. Through immersive virtual interfaces, operators can interact with robotic systems deployed in real-world environments, enabling remote monitoring, control, and manipulation. This can also serve as a testing ground for the development and training of robotic systems. Virtual environments allow engineers and researchers to simulate complex scenarios and interactions, accelerating the design and optimization of robotic algorithms and behaviors. By leveraging the immersive nature of the metaverse, robotics developers can refine their systems safely and cost-effectively before deploying them in the physical world. Such utilization of virtual environments can enhance human-machine interfaces and automation, thereby shaping the future of robotics in the digital age [6, 7, 17].

In our research, we developed a system to lay the groundwork for new metaverse infrastructure by integrating gamified virtualization into the world of robotics. We have explored the creation of a new asset digital twin infrastructure for the metaverse, with a focus on dynamic 3D animation for various use cases including immersive experiences such as board games. Utilizing a combination of research, experimentation, and real-world case studies, we aimed to provide insights, best practices, and practical guidance for organizations and professionals building metaverse projects. With this innovative exploration of asset digital twins and 3D animation, we hope to contribute to the advancement of the metaverse within the framework of digital interaction and creativity.

This project contributes to the growing field of asset digital twin technology, providing a promising prototype of a framework for the enhancement of solutions in various fields including manufacturing and security. The key contributions of the present study are as follows:

- We present the underlying inspiration for customizing and adopting digital twins in various industrial business domains ranging from games to education.
- A novel asset digital twin framework is introduced and a sample use case of that architecture is explored to confirm the realization and verification of the framework. The framework consists of 3 components: intelligent logic, a transmission file that holds motion features, and an avatar with motion primitives.
- We create a new baseline approach to be used in various case studies for pilot projects with the creation of software materials for Industry 4.0 use cases. We accordingly present a case study of a board game, peg solitaire, which validates the proposed framework.
- We adapt the solution tree calculated by the breadth-first search (BFS) algorithm in previous research to extract motion features to drive the animation.
- Finally, the modularity and effectiveness of our asset digital twin framework are experimentally validated with a board game use case and an illustrative scenario via extensive experiments.

The remainder of this study is presented as follows: Section II briefly describes related works on the existing approaches and offers a review of the literature on metaverse applications, asset digital twins, domain-specific simulation and animation with the Unity game engine, 3D character animation development, and the BFS algorithm. Section III presents the end-to-end framework of the proposed asset digital twin approach, including the methodologies employed in this work, a visualization of the pipeline, and design details of the architecture. Section VI presents the case study and relevant techniques with a description of the environment, the reconstruction phase, the implementation phase, and the experimental results across the entire system. The effectiveness of the proposed application is also evaluated. Finally, Section V concludes the study, summarizing the findings and suggestions for potential future work in this field.

## II. RELATED WORK

This section presents previous studies by researchers utilizing various algorithms and methodologies for the development of digital twin and character animation systems.

### A. METAVERSE APPLICATIONS

In recent years, the concept of the metaverse has acquired significant importance, presenting novel opportunities for the virtual prototyping of physical assets by offering immersive virtual environments. In [13], the authors designed a virtual reality system for metaverse scenarios involving human-computer interactions. The research presented in [18] contributes to the creation of realistic avatars with optimized animation. Immersive applications have also been applied to educational fields, as presented in [24, 25], by implementing multilayer or theoretical frameworks for an educational metaverse. The prototype of a new dance training system was implemented in the research presented in [26].

The new focus of researchers in more recent years has been the application of metaverse opportunities to real-life scenarios in various industries. In [43], the topic of smart manufacturing is addressed. In [45], a metaverse-based evacuation system for an educational building is explored. Other use cases of metaverse applications in education are discussed in [51, 56], while smart-city applications are addressed in [52], potential healthcare applications in [46, 27], industrial use cases in [57], and intelligent driving cases in [58]. In [44, 48, 49, 50, 53, 54, 55], the concept of the metaverse is defined in general, wide-ranging industrial usage scenarios are discussed, and metaverse architectures are presented.

### B. ASSET DIGITAL TWINS

Digital twins are digital models that represent virtual replicas of physical objects, processes, or systems to reproduce their physical behaviors by integrating sensor data, historical data, and simulation models. They are widely used for their advantages in predictive maintenance, performance optimization, prototype development, and decision support in various industries.

Asset digital twins are special types of digital twins focusing on replicating individual assets within a larger system. Asset digital twins are applicable in diverse sectors including manufacturing, transportation, education, and healthcare.

In [3], the authors aimed to present a showcase of the use of digital twins in the manufacturing industry with various technologies including the Unity engine. The research presented in [4] produced a social app as an attempt at a metaverse application in the context of digital twins using the Unity 3D engine and Microsoft Structured Query Language (SQL) Server. In [5], research was conducted for a motion capture system to allow the concept of digital twins to be used in Industry 4.0 scenarios. The research in [6] produced a prototype of human operators in line with the concept of digital twins. In [10], a sample architecture was demonstrated for connected vehicles and pedestrians as a digital twin application. In [11], the fundamentals of XR technologies in Industry 4.0 were presented and digital twins were explained as a replica of physical objects in the digital environment. In [12], involving research related to digital twins, the authors established an end-to-end pipeline from two-dimensional (2D) to 3D reconstruction based on deep learning techniques. Digital twin applications also provide significant benefits in robotics with the potential advantage of supporting the development of Automation Studio and Unity technologies [14]. Researchers proposed a digital twin diesel generator system to be used for educational purposes in [15]. The research described in [16] aimed to identify key technologies for developing digital twins in the construction domain. The concept of human digital twins was proposed for use in smart manufacturing systems in [17]. On the other hand, researchers explained the challenges and opportunities of digital twin technology in the setting of sports events in [19]. In [29], the authors introduced a digital twin-based framework in the Unity environment for additive manufacturing in the domain of smart factories. Researchers aimed to demonstrate digital twin technologies with educational case studies in [30]. Many digital twin examples for use in Industry 4.0 have been implemented in various studies, such as [31, 32, 33, 34].

### C. DOMAIN-SPECIFIC SIMULATION and ANIMATION with the UNITY ENGINE

The Unity engine provides multiplatform and cross-platform tools and realistic virtual environments for implementing domain-specific simulations and animations [35, 36, 37, 41]. A simulation in the robotic domain was developed with the Unity engine to visualize the real working environment of robots in [7]. Researchers proposed a novel simulation environment allowing physics simulations for multiphysics interactions, namely the RFUniverse, with Unity while enabling Python for robot manipulation in [8]. Simulation software was developed using the Unity platform to visualize autonomous mobile robots in logistic scenarios in [27]. Additionally, a sign language reproduction system with avatars was established using the Unity engine in [36].

### D. 3D CHARACTER ANIMATION DEVELOPMENT

3D character animation development is a quick and practical approach for obtaining realism and engagement in virtual environments, providing effective simulations of real-world scenarios. Researchers proposed a generative model for character animation by formulating the motion style transfer of tasks and motion generation in [9]. In [20], a virtual patient avatar was proposed to perform care and nursing training tasks in the health domain. The authors of [28] applied a novel method of virtual character animations using Unreal Engine 4 for human pose reconstruction by retargeting RGB video streams as a lower-cost solution.

### E. BREADTH-FIRST SEARCH (BFS) ALGORITHM

The BFS algorithm is a fundamental algorithm for data structures, facilitating applications of asset digital twins for tasks such as pathfinding in animations or generating game solutions in board games. In [1], the authors proposed an effective algorithm for mining top-k high average-utility items by employing the BFS strategy with pruning techniques. The authors of [2] discussed finding a path for the Labyrinth game by employing the BFS algorithm.

From the literature specifically addressing digital twins, it appears to be possible to control a real production system or robotic system in industry by using its digital twin. The primary goal of the present study is to establish a practical framework encompassing the intelligence component with a computational layer and animation for the creation of asset digital twins that are both simple and complex. To achieve this, we designed an easy-to-follow board game use case that utilizes readily available tools and lower-cost techniques. We intend to apply this use case as a starting point for developing various asset digital twins for Industry 4.0. Additionally, our framework can be further customized to create diverse educational materials.

## III. DIGITAL TWIN FRAMEWORK

Digital twin technology is the backbone of the industrial metaverse, connecting the real world to the virtual environment by creating both digital and visual copies of an entity. A digital twin is not merely a simple replica of a system or product; it is also applicable to various stages of industrial processes such as manufacturing, smart automation, and predictive analytics. There are four main digital twin types including component twins, asset digital twins (or product digital twins), system twins, and process digital twins. In this study, a novel asset digital twin framework for animating an avatar based on a specific scenario is proposed. Usually, an asset digital twin consists of several components for modeling complex assets like engines or production systems. It ensures the analysis of the integration and interaction of each part in a whole solution. Thus, it detects potential improvement cases and the performance and efficiency of the system.

### A. MOTION PRIMITIVES AS ATOMIC ACTIONS

Motion primitives are essential building blocks for automating motions in motion planning and control algorithms, particularly in robotics, autonomous driving, industrial automation, video games, and film. They correspond to straightforward and basic movements as a set of discrete precomputed motions that an actor (e.g., a robot, a vehicle, or an avatar) will perform to accomplish a certain movement or sophisticated tasks requiring sequences of actions [22-23, 38-39-40]. For example, in a scenario that involves kung-fu fighting, each motion sequence corresponds to a motion primitive, such as a kick or punch [21]. These primitives are frequently made to be modular, reusable, and easily joined to create more complicated motions. The four key features and attributes of motion primitives can be listed as follows: composition (i.e., the sequences of actions), reusability, parameterization (i.e., parameters such as speed, direction, duration, or joint angles), and simplicity (i.e., elementary motions such as moving forward, turning, or gripping an object).

An atomic action is an essential unit of motion or activity that is a single, distinct, indivisible activity or occurrence within a video sequence in the context of computer vision and video processing. When analyzing, examining, or annotating movies, this concept is frequently applied, particularly for activity recognition, motion tracking, or action detection. Basic building pieces known as atomic actions are used to depict specific motions, gestures, or events that take place inside video frames. Based on the video's context, atomic actions can encompass a wide range of motions and activities. Some instances of atomic actions include walking, running, sitting, standing, lying down, falling, picking up an object, dropping an object, and moving an object.

Motion primitives serve as foundational components of 3D animation and simulation. They are employed in specific domains as atomic actions to streamline the animation process, improve productivity, and offer flexibility for creative experimentation and refinement. Motion primitives are basic precalculated patterns of motions such as walking, running, jumping, gesturing, or any other fundamental movement. These primitives are usually generated by animators or motion capture systems and are kept in libraries or databases to be used again in other simulations and animations. In simulation environments, motion primitives can be used to mimic the motion and behavior of various entities such as cars, robots, animals, humans, or particles.

### B. ARCHITECTURE

The proposed asset digital twin framework architecture is shown in Fig. 1 and Fig. 2. The framework illustrated in Fig. 1 consists of three main parts, with a smart computing part, a digital 3D animation part, and a data transmission file between those two modules. The user interface is represented by the yellow rectangle on the left, entailing a textbox to enter the score selected by the user and an animated video as output. The elements inside the gray rectangle on the top are the digital replica of the animator and other simulation objects such as boards and pegs inside the Unity engine. The virtual environment contains a motion primitive database (DB), a coding layer to control animation via C#, and the generated animation. Motion primitive parameters in Unity to generate an action sequence for an avatar are Speed, Throw, and Dance. The yellow rectangle at the bottom of Figure 1 represents the transition file containing motion attributes from the DB, which were previously calculated and saved. As shown in Figure 2, the data generation pipeline is formalized following a schema that specifies a batch function to generate the game search tree in Java, a DB module holding solution vectors, and a .csv file for movement.

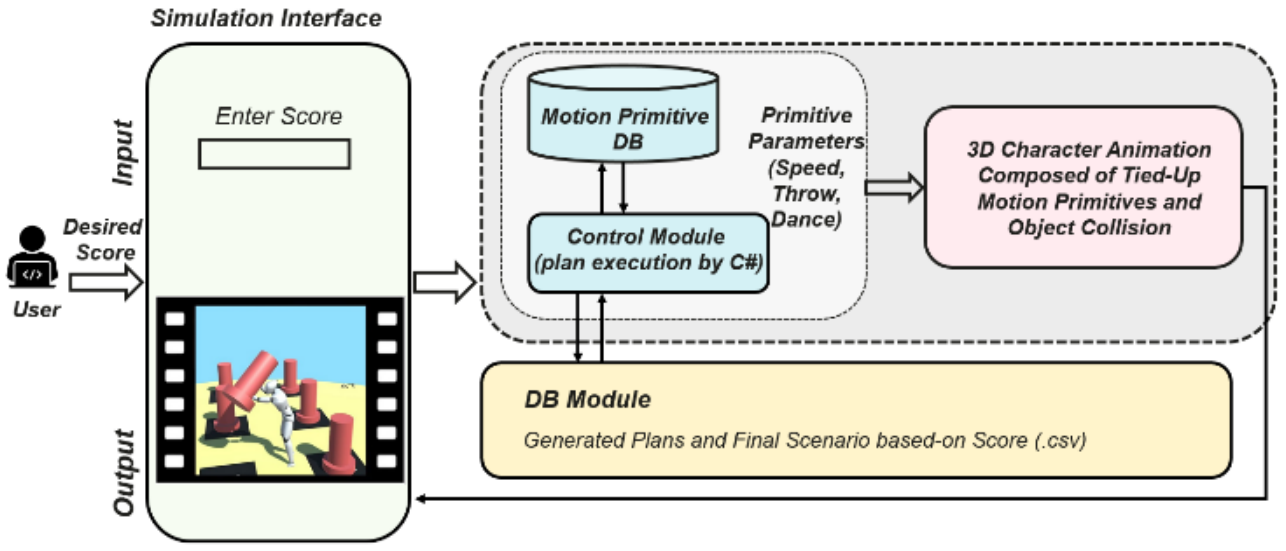

**FIGURE 1. Animation architecture overview.**

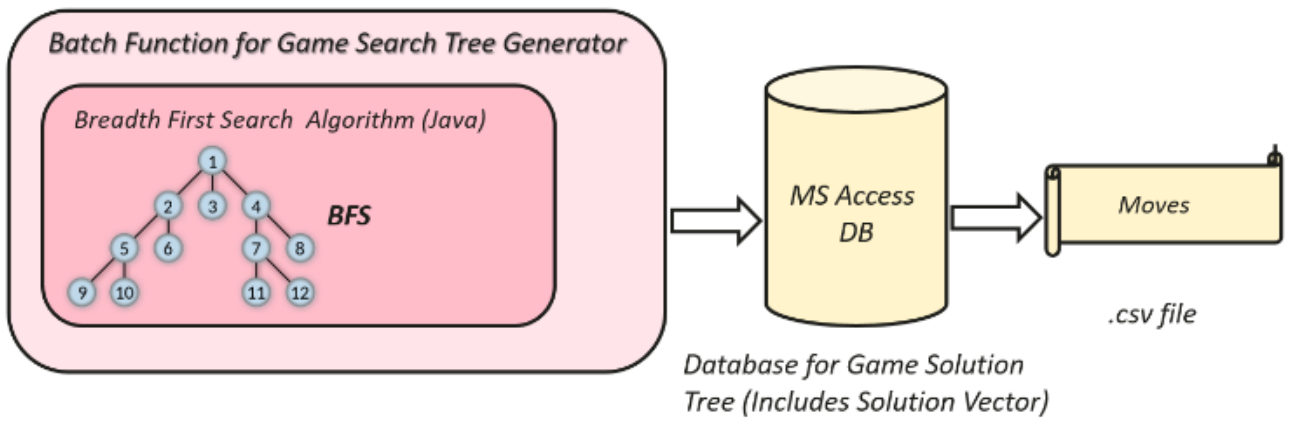

**FIGURE 2. Data generation pipeline.**

## IV. CASE STUDY: ASSET DIGITAL TWIN WITH AVATAR

In this section, a case study on an asset digital twin is presented using the proposed architecture. This case study aims to investigate how efficiently the system components work together and interact after linking the intelligent logic in the evaluation application programming interface (API) with its digital version by 3D character animation. This process allows us to obtain some choices for users with the graphical user interface (GUI), ensuring that the 3D animation is automatically performed in the digital version after selection. Our software prototype for asset digital twins was optimized with an iterative design. Iterations are performed for various use cases to upgrade the software prototype. This ensures that different types of tasks are used for iterative animation and testing, allowing us to continuously update and improve the prototype to fit various use cases and environments.

A graphical representation of interconnection in the asset digital twin system between the phases of 3D character animation development is presented in Fig. 3. Our discrete animation system functions as follows:

- Evaluation API for logic is implemented,
- Integration to the digital line is ensured, and
- The digital component interacts with users, i.e. a specific scenario is executed as 3D animation.

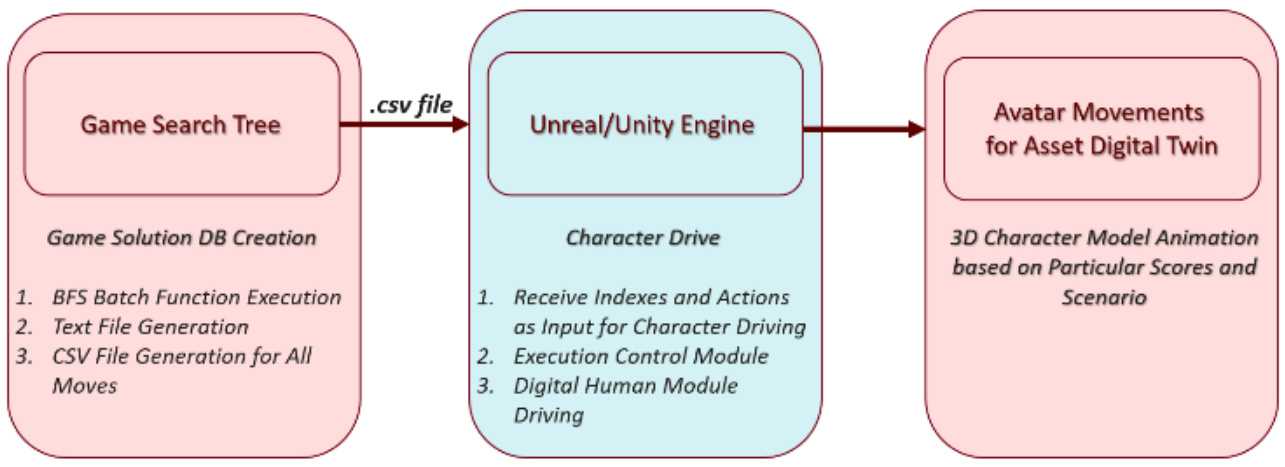

**FIGURE 3. General flow for asset digital twin system.**

In our asset digital twin system, we apply a cost-effective solution with 3D animation. In today's world, to obtain realistic animation, motion capture (Mocap) systems are widely used, especially in gaming and for metaverse solutions. Mocap systems capture all details of real actors' movements by recording them. They require expensive pieces of equipment, a motion capture studio, and highly skilled actors. For our system, we created a small animation database consisting of predefined fixed movements and chose the 3D animation technique. We created an entire animation by integrating the smaller animations according to a scenario within a script that we had evaluated.

### A. INVESTIGATIONAL TASK DESCRIPTION

The task at hand focuses on establishing an asset digital twin infrastructure integrated with 3D animation capabilities to enhance various use cases, particularly in robotics and industrial automation, and also including board games. The primary objective is to establish a versatile framework that enables the utilization of virtual representations of assets with components ranging from digital data creation to 3D characters and environmental elements. This infrastructure for asset digital twins will serve as a foundation for implementing a wide range of applications across different industries, providing a robust platform for enriched visual and interactive experiences.

The task encompasses several key components, including data generation for the game search tree, a user interface for choosing a specific game score, a .csv file for the integration, and 3D animation creation for the stated scenario, namely the board game asset of peg solitaire. Additionally, it ensures development to facilitate the creation of asset digital twins with seamless integration into product pipelines. In this work, iterative prototyping, testing, and validation of the asset digital twin infrastructure are conducted to evaluate its scalability, performance, and compatibility with specific use cases such as board games.

Peg solitaire is a classic single-player board game enjoyed by players of all ages that requires strategic thinking and puzzle-solving skills. The objective is to remove pegs from the game board by jumping them over adjacent pegs until only one peg remains. Initially, there are 33 empty holes on the game board, and 32 pegs are placed in the holes, leaving one hole empty in the center of the board. The empty hole on the game board is the starting point of the game. Pegs can be jumped left, right, down, and up, but not diagonally. A peg is moved to the empty hole immediately behind it by jumping it over another peg located to the right, left, down, or up, and the peg that has been jumped over is removed from the board. The goal of the game is to leave the fewest possible pegs at the point when no peg can be jumped over to the left, right, down, or up according to the game's rules. The score is evaluated according to the remaining number of pegs. The best score is 1, achieved when only one peg remains on the game board.

### B. ALGORITHM and DATA STRUCTURE

**Breadth-First Search (BFS or Level Order Traversal)** The BFS algorithm is used to traverse tree data structures in which all nodes at the same depth are explored before traversing to the next depth level. Generally, a queue is used to hold and track the kept tree nodes within one connected component. The traversal order of the BFS is illustrated in Fig. 4. This algorithm has many potential use cases for solving various problems in real-life scenarios, such as game search trees for chess, pathfinding with virtual maps, GPS navigation systems, robotics (e.g., robotic path planning, motion planning for mobile robots, or delivery robots), network security, and social networking.

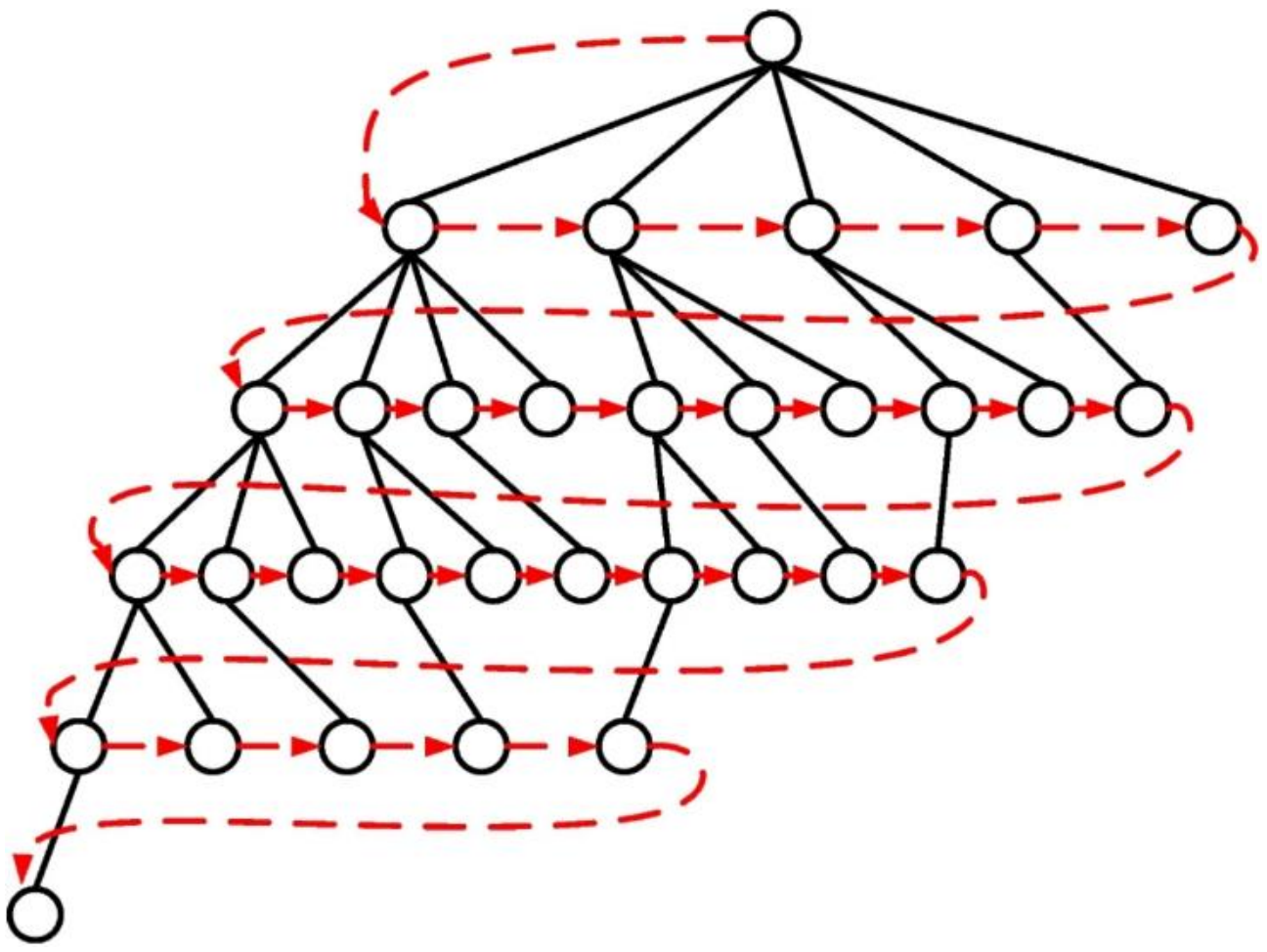

**FIGURE 4. Illustration of the tree structure and node order in the queue for the breadth-first search algorithm [1].**

The BFS algorithm is a popular graph traversal algorithm, offering distinct advantages and disadvantages depending on the problem statement. This algorithm blindly searches every node in a tree by expanding all nodes without any heuristic approach. Thus, the main advantage of the BFS is guaranteeing that a solution node or path will be found if it exists. For problem statements with more than one solution, it can find the minimal solution. In other words, the other key advantage of the BFS is its ability to find the shortest path. Other advantages are the low storage requirements, linearity, and ease of programming. However, the significant amount of memory required by the BFS is a disadvantage, and so it is not suitable for use in large-scale problems with larger trees or decision-making trees (e.g., puzzles) due to its exhaustive exploration of all solutions. This algorithm is particularly effective in scenarios that require finding the optimal path, as is the case for GPS systems. The BFS employs a first-in, first-out (FIFO) queue structure to mitigate memory constraints. To evaluate the performance and scalability of the BFS, time complexity is decisive. The time complexity of the BFS can be expressed as O(V+E) with an adjacency list representation, where O represents the upper bound of the time complexity, V denotes the number of vertices in the graph, and E signifies the number of edges in the graph [59].

Due to the performance concerns noted above, it may be necessary to optimize the BFS algorithm. Some suggested optimization methods are using queues instead of recursion to store the next nodes to be visited, performing parallel calculations, and prioritizing nodes instead of visiting all nodes in a certain order. The best optimization technique to be used depends on the particular problem statement and data type. While the BFS is effective for some types of problems, depth-first searches or A* searches may be more efficient for others. In this study, to achieve optimum performance and effective results, we store the queues in a database while creating the tree and apply a pruning method for the tree by making a selection among the nodes.

In Algorithm 1, the pseudocode of the customized BFS is provided, which is applied to calculate the game solution tree based on the rules. It includes two main functions: a main process and a function called once per node, respectively.

**Algorithm 1: Generate and save the solution tree to the DB via customized BFS**

Main process to generate solution tree in database via BFS
**Input:** pDB(pointer), pRootData(pointer)
**Output:** Search tree in database

```
function Customized_BFS(pDB, pRootData)
    #input: Data, Level, Number output: N/A
    InsertRecord_DB(pRootData, 1, 1)
    SET moveCount to 1
    FOREACH row for peg on the board // 1..31
    #input: Level, output: Data, Number
        ReadRecord_DB(row, vData, vNumber)
        vLevel = row+1
        GenerateChildren(vData, vLevel, vNumber);
        INCREMENT moveCount
    ENDFOR
end function
```

Function is called once per node
**Input:** pData, pLevel, pNumber
**Output:** Solution nodes of each node as a queue structure in tree in database

```
function GenerateChildren(pData, pLevel, pNumber)
    vIsSymmetricData_For_X = IsSymmetric_X(pData);
    vIsSymmetricData_For_Y = IsSymmetric_Y(pData);
    #input: Level, Number, output: ChildCount for Level
    GetLevelChildCount_DB(vLevelChildCount, vLevel,
vNumber)

    SET moveCount to 1
    FOREACH row for holes on the board // 0..48
        vSymmetricIndex = checkSymmetricIndex(row,
    vIsSymmetricData_For_X, vIsSymmetricData_For_Y)
        #elimination count: 3000000
        IF vLevelChildCount <= 3000000 AND
    vSymmetricIndex == FALSE THEN
            #input: Data for parent, row number for move
        output: Records for children to be inserted to db
            MoveUp(vToBeInsertedData, pData, row);
            #input: Data to be inserted, Parent Level,
        Parent Number output: Records for children on db
            InsertChildenBulkDB(vToBeInsertedData,
        vLevel, vNumber)

            MoveDown(vToBeInsertedData, pData, row);
            InsertChildenBulkDB(vToBeInsertedData,
        vLevel, vNumber)

            MoveRight(vToBeInsertedData, pData, row);
            InsertChildenBulkDB(vToBeInsertedData,
        vLevel, vNumber)

            MoveLeft(vToBeInsertedData, pData, row);
            InsertChildenBulkDB(vToBeInsertedData,
        vLevel, vNumber)
        END IF
        INCREMENT moveCount
    ENDFOR
end function
```

## C. DESIGN AND IMPLEMENTATION WITH THE UNITY ECOSYSTEM

This section provides a brief overview of the design and implementation of 3D character animation with customizable 3D avatars in the Unity ecosystem. A short description of relevant Unity concepts is also provided to highlight the key advantages that led to their choice.

**Unity:** Unity is a cross-platform graphic environment for the development of three-dimensional (3D) and two-dimensional (2D) animations and simulations. The engine is useful in various industries including the automotive industry, manufacturing, construction, film, video gaming, and engineering. In the present study, it is used as the main environment for avatar animation, connecting with intelligent logic outside of the platform via a .csv file.

**Rigging:** Rigging is one of the most important procedures in developing an avatar. It ensures the creation of a skeleton that consists of a bone hierarchy. Animation rigging also allows the user to create and organize various constraints based on a logic described in the C# animation API to implement requirements related to animation.

**Inverse Kinematics (IK):** IK is an essential process for computer animation and robotics. It entails the use of mathematical calculations to determine variable joint parameters of the motion of a character or a robot, or, in other words, the end effector, to reach a target position. In contrast, forward kinematics (FK) entails the use of mathematics to compute the position of a character or a robot from specific values for the joint parameters.

**Motion Interpolation:** Motion interpolation is a very common method used in computer animation to fill in values between frames of two keyframes and smooth the appearance of a motion. Unity provides various functions for interpolation by estimating values between two data points, ensuring animations with smoother and more realistic movement.

With the Unity Editor, avatar construction is undertaken with a character model, skeleton, and bone mapping as shown in Fig. 5. Bone mapping in the Unity environment for the avatar's head, left hand, and right hand is shown in Fig. 6.

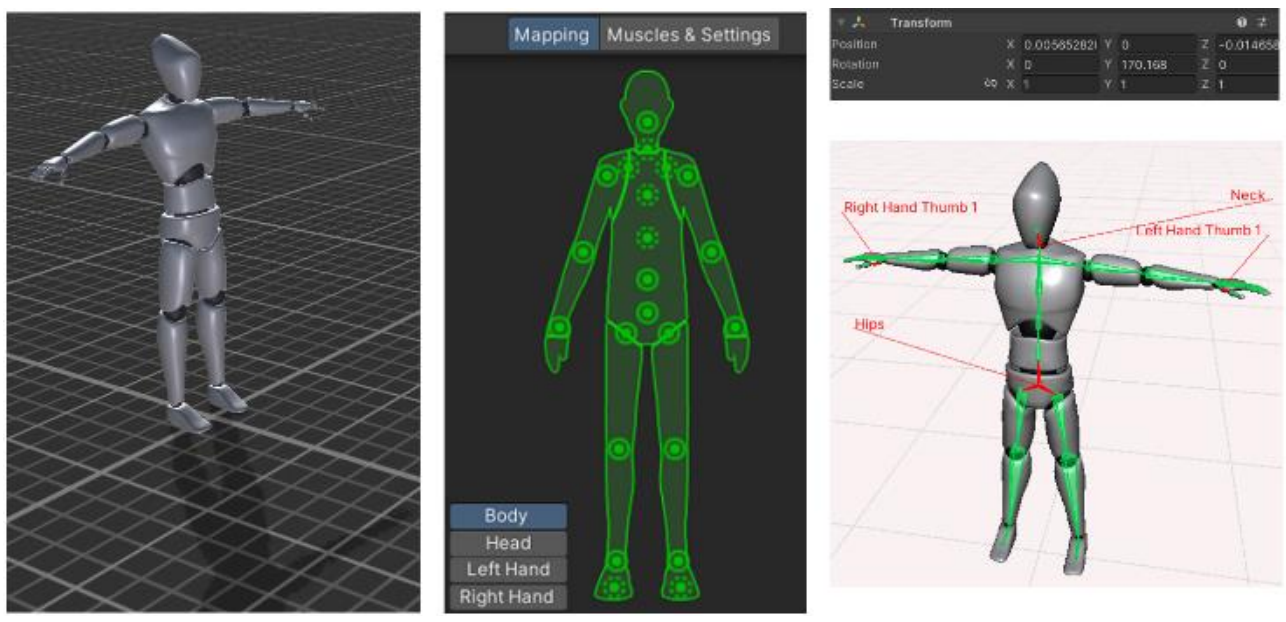

**FIGURE 5.** **Character model, skeleton, and bone mapping in the Unity environment.**

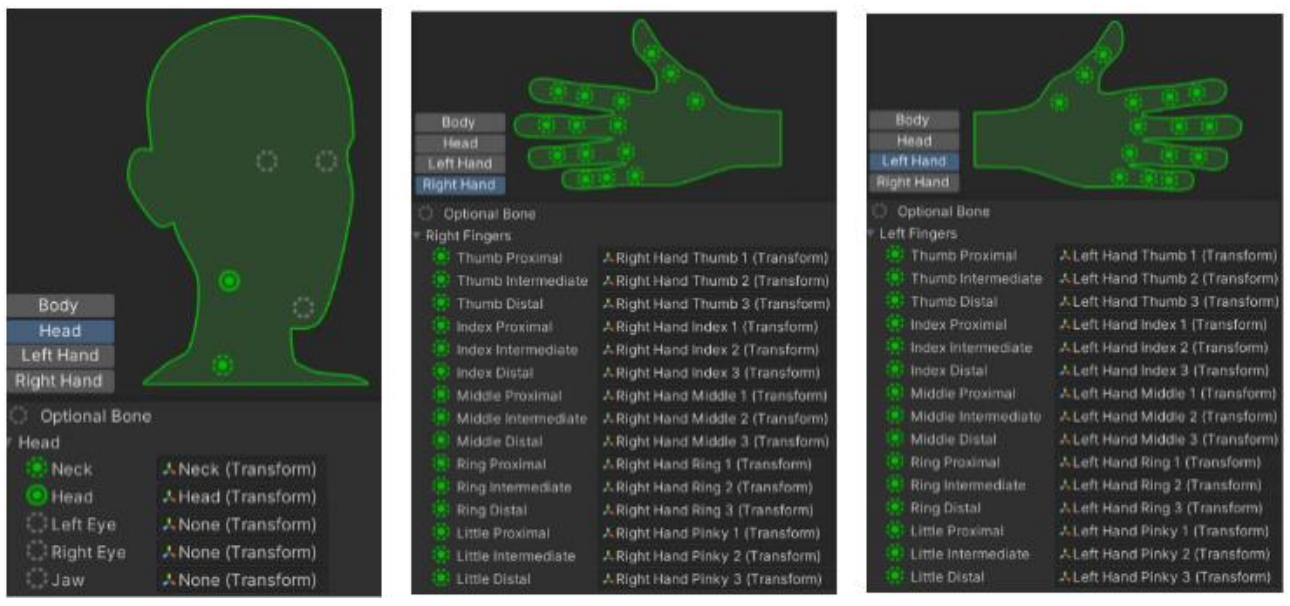

**FIGURE 6.** **Bone mapping for the avatar's head, left hand, and right hand in the Unity environment.**

Unity provides an animation tree as a layered system to simplify animation. Thus, it assists users in integrating animation clips. With a given set of parameters, transitions between states are provided. As shown in Fig. 7, the animation tree in this project has three parameters, Speed, Throw, and Dance, and they ensure control of the animation flow in a sequence within the control logic in C#. Table 1 lists the conditions for state transitions between four types of status: Idle, RunForward, MiningLoop, and Dance.

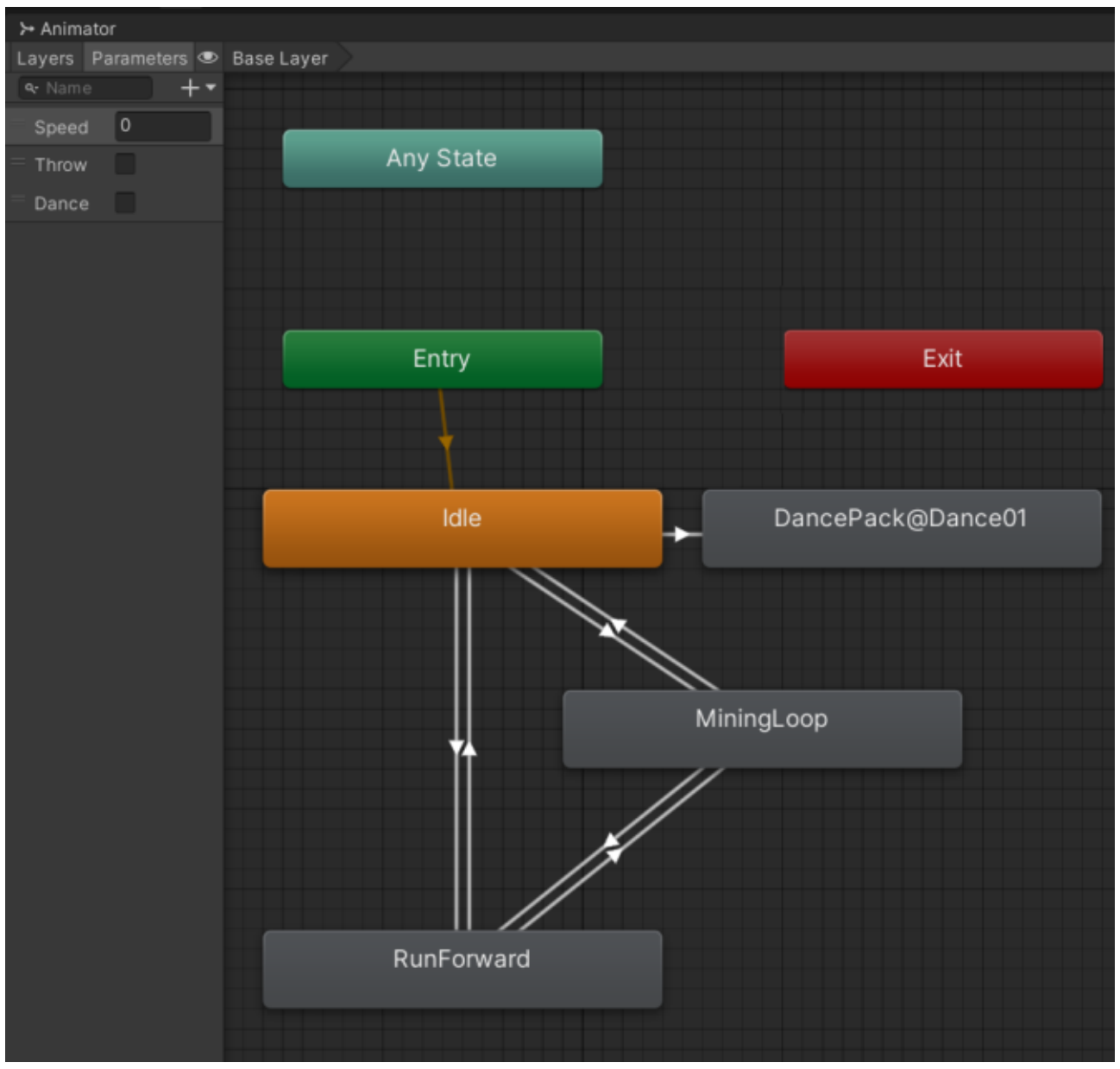

**FIGURE 7.** **Animation tree with three parameters (Speed, Throw, and Dance) to control the animation flow in a sequence.**

TABLE I

CONDITIONS FOR STATE TRANSITION

| Current Status | Next Status | Conditions |
|---|---|---|
| Idle | RunForward | IF Speed Greater Than 0.4 |
| Idle | MiningLoop | IF Throw Is True |
| Idle | Dance | IF Dance Is 1 |
| RunForward | Idle | IF Speed Less than 0.2 |
| RunForward | MiningLoop | IF Throw Is True |
| MiningLoop | RunForward | IF Throw Is False |
| MiningLoop | Idle | IF Throw Is False |

Fig. 8 presents sample sequential frames of video clips for each state, corresponding to motion primitives of 3D characters at the same time. These are predefined actions to be used to generate the proposed animation logically.

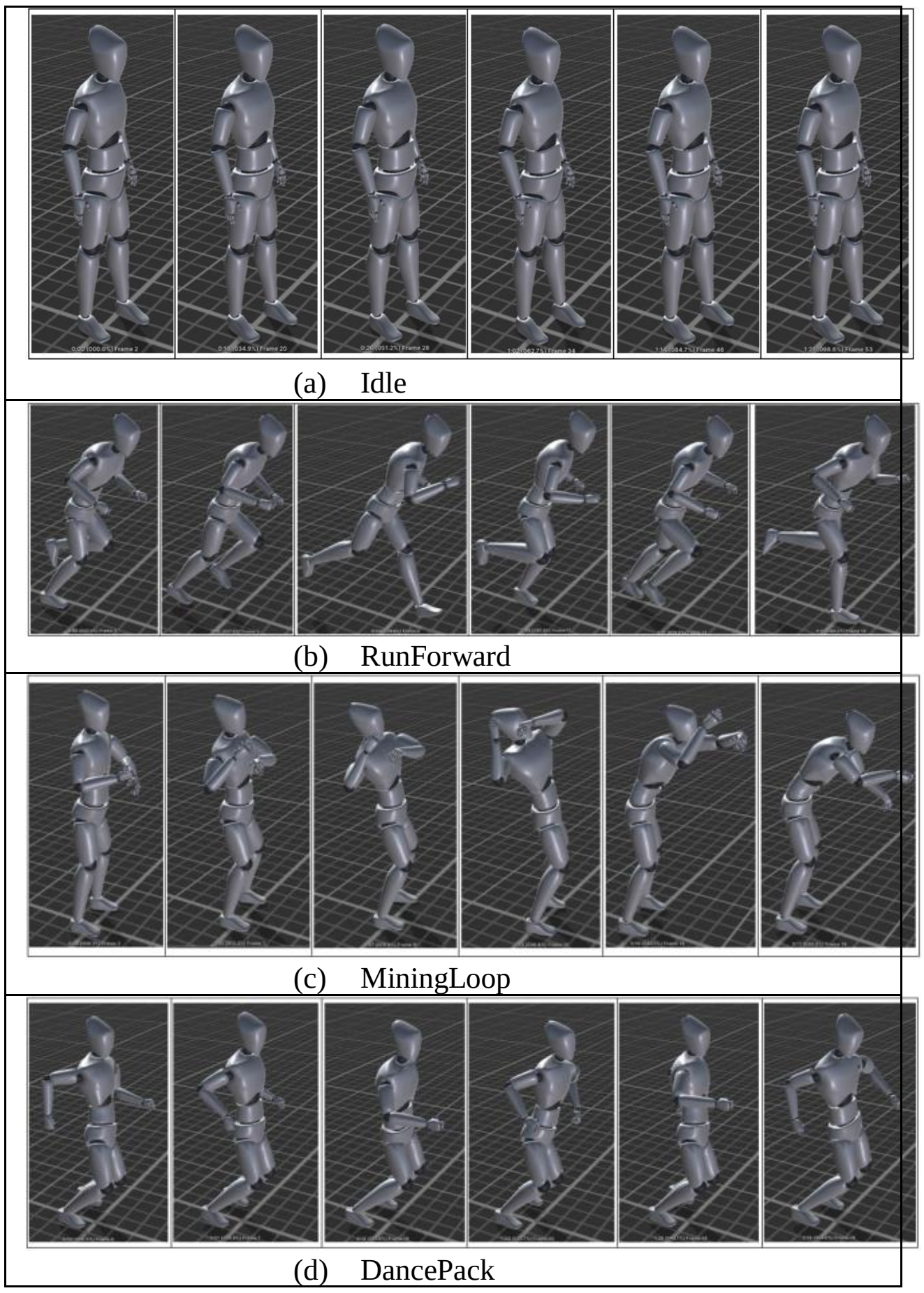

**FIGURE 8.** **Short sequence results for the avatar's poses in the Idle (a), RunForward (b), MiningLoop (c), and DancePack (d) animations produced by the proposed solution.**

The procedures for evaluating trajectories and driving 3D characters are presented in the pseudocodes of Algorithm 2. In other words, these are the control algorithms of the avatar, used to generate a list of actions and object combinations. They contain a main process and a function to be called once per frame. This allows the generation of 3D character animation for a specific scenario, in our case based on predefined movement.

**Algorithm 2: Generate 3D character animation**

**Input:** pFilePath, pScore
**Output:** Avatar Animation

```
function AnimationMainProcess(pFilePath, pScore)
    ReadCsvFile(pFilePath)
    line = ReadLine()
    WHILE line is not empty DO
        #input: line, output: game_id, game_score,
    move_index_1, move_index_2, move_index_3
        ParseLine(line, game_id, game_score, move_index_1,
    move_index_2, move_index_3)
        ChoosePeg(move_index_1)
        DestroyPeg(move_index_3)
        line = ReadLine()
    ENDWHILE
    ShowScore()
    #stop game with dance
    StopGame()
end function
```

Function is called once per frame
**Input:** target_index, move_index_1, move_index_2
**Output:** Character Movement

```
function ProcessFrameCharacterMovement(pTargetIndex,
pMoveIndex1, pMoveIndex2)
    IF pTargetIndex is not null THEN
        CalculateDirection(pTargetIndex)
        MoveToDirection(pTargetIndex)
        SetAnimationSpeed()
        IF Speed Greater than 0.4 THEN
            SetAnimationState("RunForward")
        END IF
        IF Speed Less than 0.2 THEN
            SetAnimationState("IdleMode")
        END IF
        near_objects =CheckNearbyObjects()
        FOREACH object IN near_objects
            IF object == all_objects(pTargetIndex) AND
        pTargetIndex == pMoveIndex1 THEN
                GrabObject(pTarget_Index)
            ELSIF object == all_objects(pTargetIndex)
        AND pTargetIndex == pMoveIndex2 THEN
                ReleaseObject(pTargetIndex)
            ELSE
                RotateRight()
            END IF
        ENDFOR
    ELSE
        SetAnimationState("IdleMode")
    END IF
end function
```

### D. EXPERIMENTAL RESULTS

We implemented and tested our framework end-to-end to demonstrate the effectiveness of the pipeline for a board game as an illustrative example. We divide the experimental results into three subsections: data generation, the integration file, and avatar simulation via the reconstruction of the game in the virtual environment. As an execution and an exploratory attempt for various experimental trials, we define a GUI to choose a solution from the game search tree.

#### 1) DATA GENERATION FOR GAME SOLUTION TREE

In this study, the BFS algorithm was employed to optimally assist in generating a game search tree for peg solitaire. The algorithm uses a queue and DB to calculate and keep the solution space for the problem. The first step is creating a search space for the board game. The moves in the game solution tree were calculated with a procedure written in Java using the BFS algorithm and the queue data structure, and then the game tree was saved in the DB. An elimination strategy was developed to accelerate the search algorithm and the final scores were calculated within a reasonable time. With the procedure that creates the solution tree, 8585713 nodes were generated and held in the DB, and 37397 solutions with various scores from 1 to 26 were calculated. Sample scores from 1 to 26 in the DB were selected and the game moves were saved to a text file. The game moves in that text file were then transferred to a .csv file and read with the C# function in Unity. This file contained indexes of pegs to be jumped, moved, and removed in order, and so the values ensured avatar position control. Java and the Microsoft Access DB were used for the programming stages. Table 2 shows sample results of the data patterns from the DB after calculating the game search tree according to tree level order and queue structure. In Table 3, solution tree nodes are given with calculations and iterations.

TABLE II

TABLES FROM ACCESS DB

| Table_ Solo_Test_ Result | Table_ Solo_Test_ Moves | Table_ Solo_Test_ Moves_Temp | |
|---|---|---|---|
| Level | Number | First_Peg_1 | First_Peg_2 |
| 32 | 2 | 10 | 15 |
| 32 | 1 | 10 | 15 |
| 32 | 93 | 10 | 15 |

TABLE III

GAME SEARCH TREE CALCULATION

| Table_ Solo_Test_ Result | | Table_Solo_Test_Moves | | Table_ Moves_ Temp | | |
|---|---|---|---|---|---|---|
| ID | c_level | c_number | Data | Evaluation | Parent_Level | Parent_Number |
| 8585752 | 32 | 1 | 9900099990009900000000000000 0000000000990009999901099 | 2 | 31 | 22 |
| 8585727 | 31 | 5 | 9900099990009900000000000000 0000000000990009999910199 | 2 | 30 | 28 |
| 8584366 | 2 | 1457 | 9900099990009900000000000000 0000000000991119999901099 | 2 | 28 | 4063 |

**Queue Structure.** Nodes in each level of the solution tree are kept in a queue structure in the DB. The initial state, or the root node, is saved to the queue at Level 1. The child nodes of that root node are calculated according to the game rules by the BFS algorithm and then added to the queue at Level 2. This process is repeated for each level in a loop. The number of pegs on the board is the same for the nodes of the same level in the DB. As the level number increases, the number of pegs decreases by 1. For example, at Level 1, there are 32 pegs on the board, with 1 remaining peg on the board at Level 32. Table 4 shows the data pattern in the DB to be used in the calculation of the solution tree.

TABLE IV

DATA PATTERN OF THE GAME SEARCH TREE IN THE DB

| Field Name | Data Type |
|---|---|
| Level | Number |
| Number | Number |
| Data | Text |
| Evaluate | Number |
| Parent_Level | Number |
| Parent_Number | Number |
| Move_Peg_Skip_No | Number |
| Move_Peg_Locate_No | Number |
| Move_Peg_Remove_No | Number |
| For_One_Peg | Number |
| Is_It_Score | Number |

As seen in Fig. 9, the node structure includes a level for movement number, a number for queue sequence, data for representing peg numbers on the board, and Parent_Level and Parent_Number for linking the node and the parent of the node in the tree hierarchy.

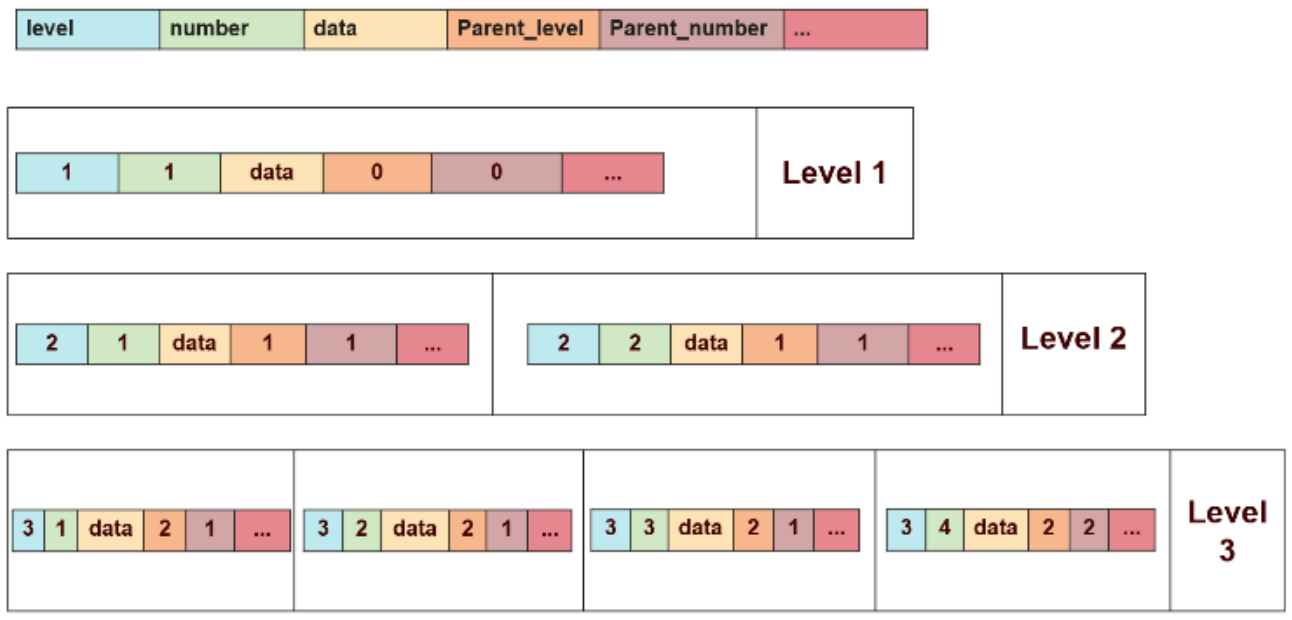

FIGURE 9. Node structure.

**BFS Algorithm.** The solution tree was created using the BFS algorithm as seen in Fig. 10. The tree consists of the following order with the BFS approach: 1, 2, 3, 4, 5, 6, 7, 8, 9, 10, 11, 12, 13, 14, 15. The solution tree is created by an iterative method; it starts from the root, and then child nodes are created by applying the rules. With each iteration, one more level is created in the tree, and so the number of iterations is equal to the height of the tree.

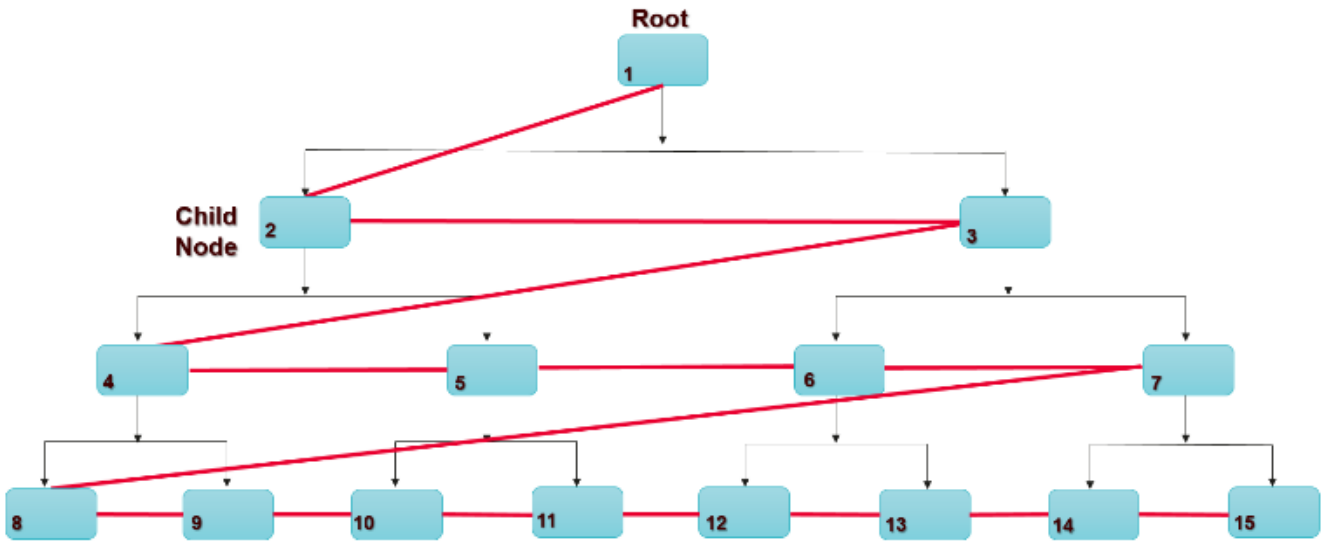

FIGURE 10. Level order tree traversal with BFS algorithm.

Table 5 shows the numbers of various solutions for each level and score. For a score of 1, two different solutions were calculated, while 26 different solutions were calculated for a score of 2. A total of 37397 solutions were calculated [42].

TABLE V

TABLE MOVES FOR GAME SEARCH TREE

| Score | Level | Total Number of Solutions |
|---|---|---|
| 1 | 32 | 2 |
| 2 | 31 | 26 |
| 3 | 30 | 186 |
| 4 | 29 | 829 |
| 5 | 28 | 2336 |
| 6 | 27 | 5448 |
| 7 | 26 | 7786 |
| 8 | 25 | 8430 |
| 9 | 24 | 5806 |
| 10 | 23 | 3551 |
| 11 | 22 | 1311 |
| 12 | 21 | 432 |
| 13 | 20 | 209 |
| 14 | 19 | 35 |
| 15 | 18 | 1 |
| 17 | 16 | 3 |
| 18 | 15 | 2 |
| 21 | 12 | 2 |
| 22 | 11 | 1 |
| 26 | 7 | 1 |
| TOTAL | | 37397 |

### 2) INTEGRATION FILE

The integration file includes the final digital pieces of information, as shown in Table 6. To obtain a score of 1 as a sample scenario, movements in a sequence are given in a .csv file. This is used to drive the motion. In our particular use case, we preferred a .csv file solution because of the precalculated game scenario. In the future, we will reevaluate this approach using the Microsoft SQL Database or Firebase to integrate the digital twin that we created into real-time application scenarios.

TABLE VI

SAMPLE MOVEMENT ATTRIBUTES IN .CSV FILE

| game_id | game_score | move_index_1 | move_index_2 | move_index_3 |
|---|---|---|---|---|
| 101 | 1 | 26 | 24 | 25 |
| 101 | 1 | 11 | 25 | 18 |
| 101 | 1 | 20 | 18 | 19 |
| 101 | 1 | 34 | 20 | 27 |
| 101 | 1 | 17 | 19 | 18 |
| 101 | 1 | 20 | 18 | 19 |
| 101 | 1 | 25 | 11 | 18 |
| 101 | 1 | 15 | 17 | 16 |
| 101 | 1 | 2 | 16 | 9 |
| 101 | 1 | 11 | 9 | 10 |
| 101 | 1 | 4 | 2 | 3 |
| 101 | 1 | 17 | 15 | 16 |
| 101 | 1 | 14 | 16 | 15 |
| 101 | 1 | 29 | 15 | 22 |
| 101 | 1 | 28 | 14 | 21 |
| 101 | 1 | 15 | 17 | 16 |
| 101 | 1 | 2 | 16 | 9 |
| 101 | 1 | 17 | 15 | 16 |
| 101 | 1 | 14 | 16 | 15 |
| 101 | 1 | 23 | 9 | 16 |
| 101 | 1 | 37 | 23 | 30 |
| 101 | 1 | 32 | 30 | 31 |
| 101 | 1 | 30 | 16 | 23 |
| 101 | 1 | 9 | 23 | 16 |
| 101 | 1 | 23 | 25 | 24 |
| 101 | 1 | 46 | 32 | 39 |
| 101 | 1 | 25 | 39 | 32 |
| 101 | 1 | 44 | 46 | 45 |
| 101 | 1 | 46 | 32 | 39 |
| 101 | 1 | 33 | 31 | 32 |
| 101 | 1 | 38 | 24 | 31 |

### 3) AVATAR SIMULATION

The primary step in the generation of asset digital twins in the virtual world for the metaverse, including the construction of the environment and the coding module, is the 3D avatar animation. Based on the asset digital twin framework and design criteria proposed in this study, we implement a prototype for a 3D character animation system compatible with multiple platforms, including PC and Android mobile applications. Unity’s superiority across comprehensive animation and simulation development ecologies prompted its selection as a virtual simulation environment for avatar animation and interaction.

The modeling process and the basic principles of the workflow of the program in this work are as follows: The process starts with the user’s action of entering the desired score on the GUI screen. First, movement attributes are selected from a file to prepare the animation scenario. The animator status then changes to “Run Forward” from “Idle” across the desired peg based on the scenario and the avatar starts running to pick up the peg. When the animator approaches the desired peg, a collision occurs between the object and the hand picking up the peg. The hand holds the peg, and so the animator continues to run with the peg in hand to the position to which it will be moved. As soon as the animator reaches the new position, the status changes to "Mining Loop" from "Run Forward," and the animator places the peg in its new position. After that, the animator status changes from "Mining Loop" to "Idle." A new movement cycle then starts and continues until the end of the game. Finally, the score is calculated, and then the game ends and the animator begins dancing.

In Fig. 11, the best frames representing movement are illustrated with the decomposition of a move on the board into the stages realized by motion primitives and the coding layer. We created animations by concatenating predefined motion clips based on the scenario in a C# coding module.

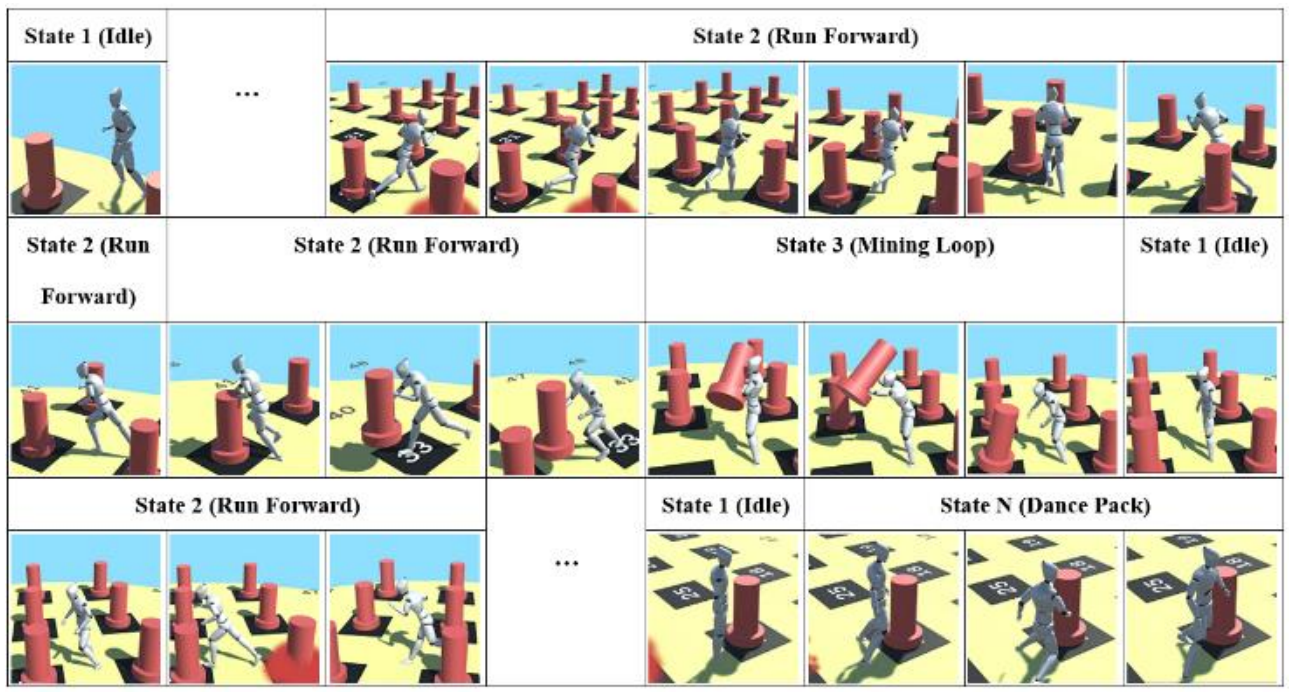

**FIGURE 11.** **Decomposition of movement on the board into stages realized by motion primitives and the coding layer.**

We conducted extensive experiments to validate the quality of the proposed asset digital twin pipeline utilized by our method, as seen in Fig. 12. The prototype proposed in this study achieves its goal of generating smooth animation based on a predefined scenario using motion primitives with a cyclic pattern. To highlight remarkable results, we extracted successive frames representing movements with motion sequences. The animation for one movement step in the game is the same and it repeats cyclically. Fig. 12, illustrating outcomes for scores of 1 and 15, shows that animation is achieved based on the chosen scenario and the full-body poses of the avatar and the appearance of the objects are satisfactory.

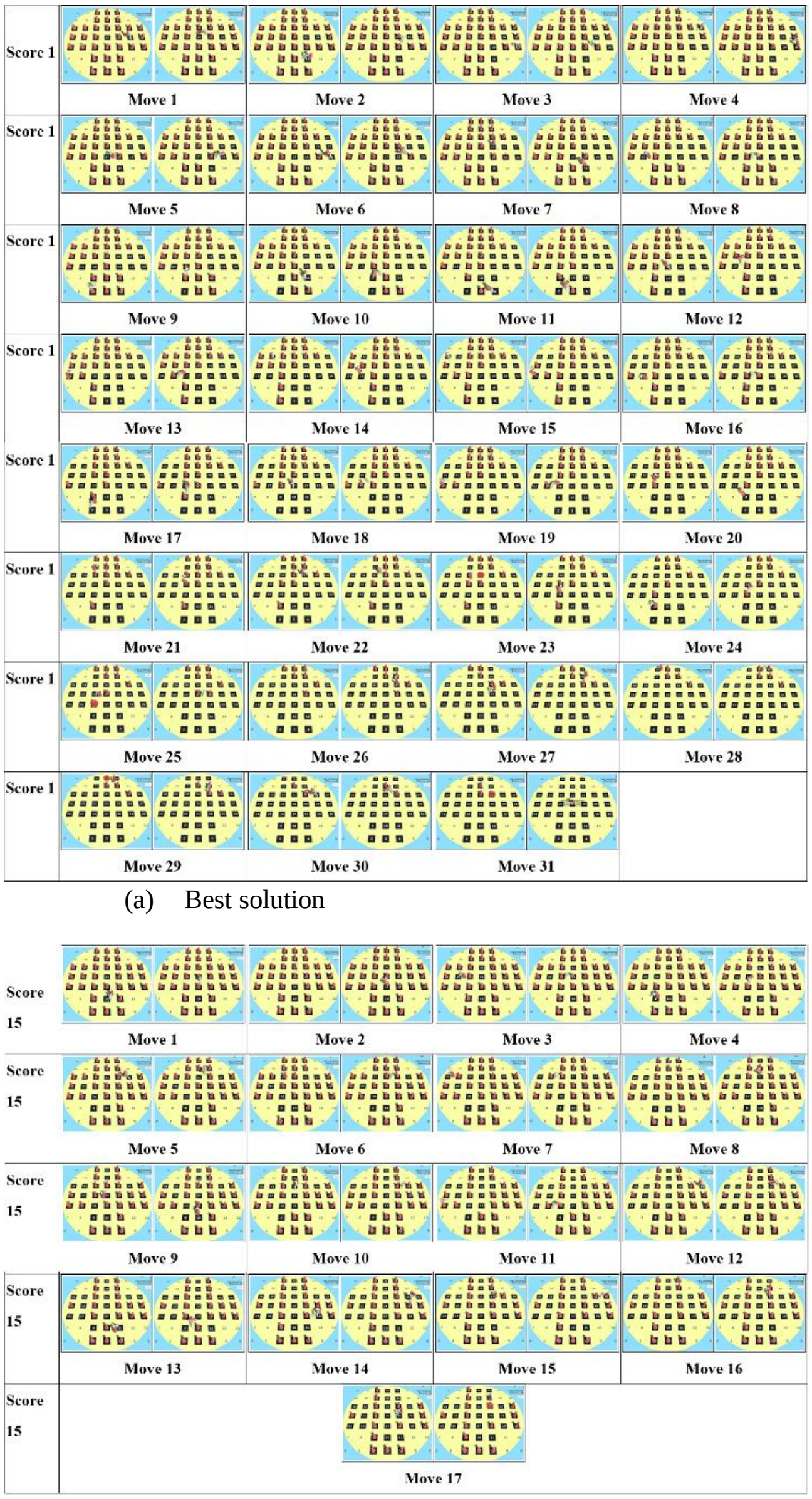

(a) Best solution

(b) Good solution

**FIGURE 12.** **Qualitative results obtained by the avatar animation.**

**Animation Performance Analysis.** Comprehensive experiments were conducted to gain further insight into our proposed framework. The performance evaluation of the game obtained based on extensive simulation results is presented here, including the measurements and performance calculations for a recording from gameplay. For the performance analysis of the animation, emphasis was placed on 60 FPS count, CPU usage, the rendering process, and memory. Before presenting the performance results for the animation of the game, however, the case study should be further explained. The total number of pegs that could be on the gameboard at one time was 32 and they appeared in order. Data recording took place on a personal computer with the Unity Profiler. Board games are more stable compared to other games for which characters are animated according to dynamic scenarios. In our case study, there is no user control. The main character of the game moves within the predefined map in a precalculated scenario, picking up and placing pegs while redundant ones are removed. Since the character follows a predefined and almost fixed path with fixed objects, the resource consumption in the game is generally always the same.

The Profiler graphs in Figure 13 show the usage of components for gameplay on a personal computer. From the CPU usage diagram, we see that the game is largely stable with rare spikes and it runs at 60 FPS, which is an optimum FPS value. Furthermore, the line for GC allocation is rarely activated in the game after the memory reaches a certain threshold. In the event that GC is active, with spikes in the graph, and then the memory is cleared, the GC allocated in the frame decreases.

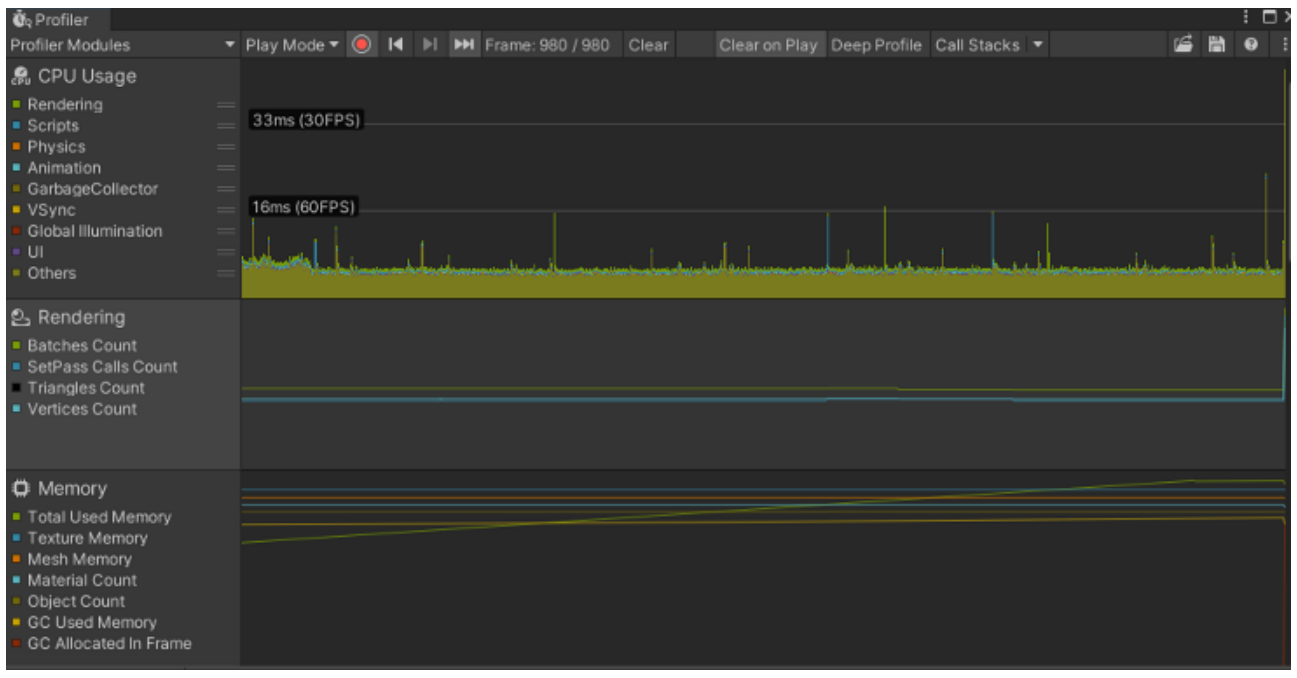

**FIGURE 13. Profiler results of game animation for gameplay.**

Table 5 summarizes the data collected for a frame with approximately 60 FPS. Most variables showed good results, such as set pass calls, draw calls, and batches. If shadows and lighting are ignored, the animation performance is even better.

TABLE V

VALUES FOR A FRAME WITH 60 FPS FOR CPU USAGE

| Variables | Value |
|---|---|
| Set Pass Calls | 72 |
| Draw Calls | 987 |
| Batches | 987 |
| Triangles | 297.1K |
| Vertices | 237.1K |
| Used Textures | 17 / 1.1 MB |
| Render Textures | 16 / 66.7 MB |
| Render Textures Changes | 14 |
| Used Buffers | 215 / 5.4 MB |
| Vertex Buffer Upload in Frame | 4 / 0.8 KB |
| Index Buffer Upload in Frame | 4 / 48 B |
| Shadow Casters | 552 |

Table 6 presents data for CPU usage with the target frame rate (15.2 ms + 1.1 ms). The target frame rate is the FPS that Unity targets for rendering the animation. For a smoother animation experience, 60 FPS would be the optimum value. For instance, if the target frame rate is 60 FPS, then the frame time is equal to 16.66 ms (1000 ms/60 FPS). In other words, it takes 16.66 ms to render each frame in Unity at 60 FPS. In our experiments, as shown in Table 6, CPU usage is 16.3 ms in total, reflecting a value of about 60 FPS.

TABLE VI

STATISTICS OF GRAPHICS FOR CPU USAGE

| Variables | Value |
|---|---|
| CPU (Main) | 15.2 ms |
| CPU (Render Thread) | 1.1 ms |
| Batches | 502 |
| Tris | 149.0K |
| Verts | 119.1K |
| Screen | 554×703 – 4.5 MB |
| Set Pass Calls | 40 |
| Shadow Casters | 279 |
| Visible Skinned Meshes | 2 |

In Table 7, memory data for CPU usage are presented. The total allocated memory is 0.90 GB and the total used memory is 242.4 MB, as seen in the table. Thus, the animation does not incur high memory consumption.

TABLE VII

MEMORY DATA FOR CPU USAGE

| | Variables | Value |
|---|---|---|
| **Memory Usage on Device** | Total Allocated | 0.90 GB |
| | Total Resident on Device | 242.4 MB |
| **Allocated Memory Distribution** | Executables & Mapped | 351.4 MB |
| | Native | 177.7 MB |
| | Managed | 10.7 MB |
| | Graphics (Estimated) | 155.6 MB |
| | Untracked | 227.8 MB |
| **Managed Heap Utilization** | Virtual Machine | 8.9 MB |
| | Empty Heap Space | 0.9 MB |
| | Objects | 0.9 MB |
| **Top Unity Object Categories** | Render Texture | 134.8 MB |
| | Mesh | 2.5 MB |
| | Shader | 1.9 MB |
| | Texture2D | 1.6 MB |
| | Audio Manager | 1.2 MB |
| | Others | 3.4 MB |

There are more than 211 objects on our screen, including 49 cells, 49 canvases, 49 texts, 32 pegs, and a total of 32 bases corresponding to each peg. Although many objects are drawn on the screen, evaluations of the animation, memory, target frame rate, and GC results confirm that the performance is good compared to animations with dynamic paths. Furthermore, our animation system achieves performance comparable to the game performance reported in [60] in terms of CPU usage, rendering, and memory usage. In future work, we will integrate our solution into mobile devices and optimize the performance by exploring basic optimization techniques for mobile devices.

## V. CONCLUSION

To demonstrate the potential of asset digital twins, we developed a data-driven avatar-based 3D animation system in Unity for a specific use case and presented all the work entailed for the completion of four conceptional stages: the generation of the game search tree, motion primitives, trajectory planning in coding logic, and animation generation. Within that framework, an asset digital twin approach incorporating programming logic and avatars was proposed and executed in the specific scenario of a selected board game. The framework consists of an algorithm, a digital world via Unity, and data transmission via a file between the intelligence logic and the digital spaces. A sample use case was established to validate the architecture of the framework and the asset digital twin system was further validated in a board game application, namely peg solitaire. This architecture combines three platforms: the BFS algorithm in Java, a solution tree saved to Access DB, and Unity for asset digital twinning with 3D characters and objects, respectively. The Java and DB platforms constitute the driving animator to animate the scenario, while the 3D character option of Unity is used to produce the animation environment for the board game. Data transmission between different platforms is provided by a file structure with attributes matching movements. To validate the effectiveness of the asset digital twin framework, a case study was conducted to animate a board game based on the proposed architecture. The case study demonstrated the benefits of the proposed data-driven 3D animation system by combining motion primitives and coding logic to drive animators within a modular structure. The presented asset digital twin framework is expected to serve as a powerful architecture for future research on digital twin technology to enhance software tools and applications in manufacturing, safety, education, health, and mobility. In future research, we will continue building on AI systems and expand our current framework to incorporate other use cases such as chess, human motion imitation, motion reconstruction with motion primitives for predefined scenarios, robotic simulations, sign language production, physiotherapy exercises, smart manufacturing, and more complex Industry 4.0 business scenarios. Furthermore, AI algorithms such as transformers, generative adversarial networks, 3D convolutional neural networks, and long short-term memory (LSTM) neural networks for the generation of motion data to drive avatars could be investigated using the proposed asset digital twin framework.

## ACKNOWLEDGMENT

The authors acknowledge the use of ChatGPT 3.5 (https://chatgpt.com/) to suggest an outline structure for some generic English sentences related to the topics of this paper.

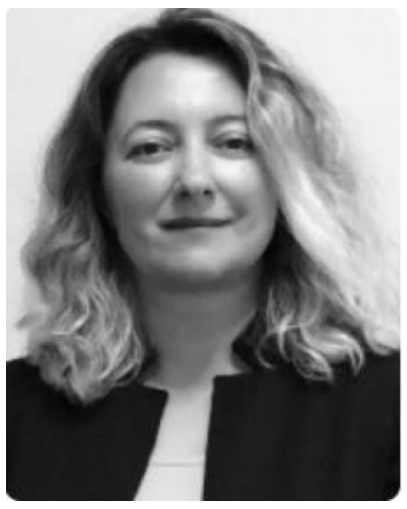

**Senem Tanberk** was born in Düzce, Türkiye, in 1976. She received a BSc degree in control and computer engineering from Istanbul Technical University (Türkiye) in 1998, MSc degree in mechatronics from Marmara University (Türkiye) in 2012, and PhD in computer engineering from Doğuş University (Türkiye) in 2020. She has also worked as an IT professional in the telecommunications, finance, and banking sectors for over 15 years in Istanbul, Türkiye. She has expertise in Telco pricing and billing, revenue and financial reports, DWH support, software design and development, and advanced PL/SQL (Oracle) optimization. She is also a Telco Oracle PL/SQL trainer. She is currently a research and innovation manager in a private company in Istanbul. Her research interests include deep learning, data science, robotic programming, 3D character animation, and simulation programming.

.

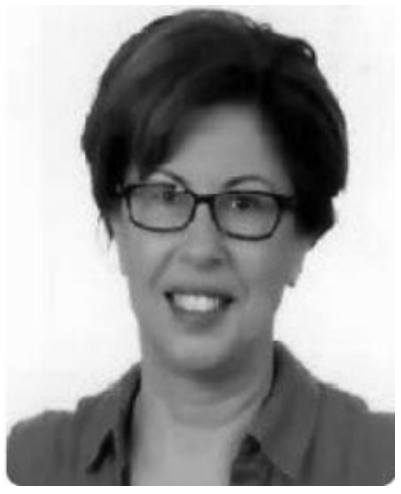

**Dilek Bilgin Tükel** was born in Adana, Türkiye, in 1965. She received a BSc degree in electrical engineering in 1987, MSc from Boğaziçi University (Türkiye) in 1990, and PhD in mechanical engineering from Katholieke Universiteit Leuven (Belgium) in 1997. She is an Assistant Professor in the Software Engineering Department of Doğuş University (Türkiye) and Research and Development Leader with Altinay Robot Technologies, Istanbul, Türkiye. Her research interests include the modeling and control of robotic systems, industrial automation systems, AR/VR, and the IoT.

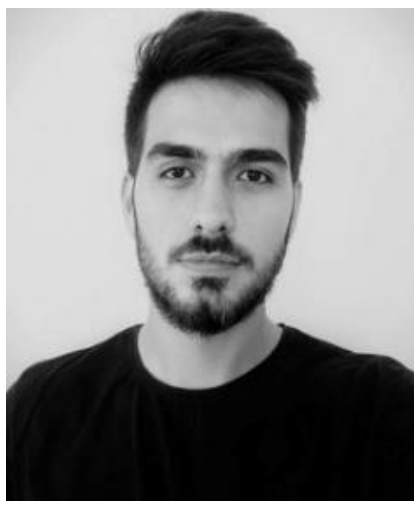

**Kadir Acar** was born in Istanbul, Türkiye, in 1999. He received a BSc degree in computer engineering from Karabük University (Türkiye) in 2022. He is a proficient user of Unity and experienced in the development of mobile games. Interested in cybersecurity, he received a cybersecurity specialist certificate from Boğaziçi University (Türkiye) and is currently working towards his MSc degree in cybersecurity at Gebze Technical University (Türkiye). He is currently employed as a software engineer at Altinay Robot Technologies, Istanbul, Türkiye. His work includes the designing of motion control systems for industrial robots.